\documentclass[12pt]{article}
\usepackage{amsmath,amssymb}
\usepackage{epsfig,epsf}
\usepackage{graphicx, subfigure}
\usepackage{graphics}
\newcommand{\be}{\begin{eqnarray}}
\newcommand{\ee}{\end{eqnarray}}
\newcommand{\nn}{\nonumber \\}

\newcommand{\p}[1]{(\ref{#1})}

\newcommand{\vecb}[1]{{\bf #1}}
\def\theequation{\arabic{section}.\arabic{equation}} 
\begin{document}

\begin{titlepage}
 \begin{center}

{\Large Vacuum structure in  $3d$ supersymmetric gauge theories}
\vspace{1cm}

A.~V. Smilga

\vspace{1cm}

{SUBATECH, Universit\'e de Nantes, \\
4 rue Alfred Kastler, BP 20722, Nantes 44307, France
\footnote{On leave of absence from ITEP, Moscow, Russia}\\
E-mail: smilga@subatech.in2p3.fr}
  \end{center}

\vspace{2cm}

\begin{abstract}
Based on a talk given at the
Pomeranchuk memorial conference in ITEP in June 2013, we review 
the vacuum dynamics in $3d$  
supersymmetric Yang-Mills-Chern-Simons theories with and without extra 
matter multiplets.
By analyzing the effective Born-Oppenheimer Hamiltonian in a small 
spatial box,
we calculate  the number of vacuum states (the Witten index)  and examine their
structure for these theories. The results are identical to those obtained 
by other methods. 
\end{abstract}

\end{titlepage}

\section{Introduction}

Probably, the best known scientific achievement of Isaak Yakovlich Pomeranchuk 
was the concept of the vacuum Regge pole that is nowadays called  the {\it pomeron}.
I did not have a chance to meet Pomeranchuk  personally --- I came to ITEP when he was already gone. 
But I heard many times from his colleagues and collaborators that Isaak Yakovlich atrributed a great 
significance to 
studying   properties of the vacuum,
 and even used to joke about an urgent need for the ITEP theory group to buy
 a powerful pump for that purpose.
 
Pomeranchuk did not know that, with the advent of supersymmetry, the issues
 of vacuum structure and vacuum counting 
would acquire a special interest. The existence of supersymmetric vacua 
(ground states of the Hamiltonian annihilated by the action of supercharges and having  zero energy) 
shows 
that supersymmetry is unbroken, 
while the absence of such states signals spontaneous breaking of supersymmetry. 
The crucial quantity to be studied in this respect is the {\it Witten index}, 
the difference between the numbers of 
bosonic and fermionic vacuum states, which also can be represented as 
\be
\label{Witind}
I \ =\ {\rm Tr} \{ (-1)^F e^{-\beta H} \} 
 \ee
where $H$ is the Hamiltonian and $F$ is the fermion charge operator. 
Due to supersymmetry, nonvacuum contributions in the trace cancel. 
It is important that the quantity \p{Witind} represents 
an {\it index}, a close relative of the Atiyah-Singer index and other topological invariants, 
which is invariant under  smooth Hamiltonian deformations. The
latter circumstance allows one to evaluate the Witten index for rather complicated theories: 
it is sufficient 
to find out a proper simplifying deformation.  

My talk  (based on three recent studies \cite{ja1,ja2,ja3} )
is devoted exactly to that. I will study the vacuum dynamics in a particular class of theories ---
supersymmetric 3-dimensional gauge theories involving the Chern-Simons term.  
 Such theories  have recently attracted  a considerable attention in view of newly
discovered dualities between certain  ${\cal N} = 8$ and ${\cal N} = 6$ versions of these theories
and the respective string theories on $AdS_4 \times S^7$ or $AdS_4 \times \mathbb C \mathbb P^3$ backgrounds 
\cite{3dualityB,3dualityA}. 
\footnote{Better known is the Maldacena duality between the 4d ${\cal N} = 4$ SYM and string theory on
$AdS_5 \times S^5$ \cite{Mald,GKP}. A nice review of this topic
 was recently published in {\it Physics-Uspekhi} \cite{Gorsky} . }
Note, however, that the field theories dual to string theories are conformal and do not involve a mass gap.
In such theories, the conventional Witten (alias, toroidal) index we are interested in here is not well defined, and 
 the proper tool to study them is the so called superconformal (alias, spherical)
index \cite{Romel,Spir}.

We calculate the index by deforming the theory,  
putting it in a small spatial box and studying the dynamics of the Hamiltonian thus obtained in the 
framework of the Born--Oppenheimer (BO) approximation. The results coincide with those obtained by other methods.

Let us discuss first  the simplest such theory, the ${\cal N} = 1$ supersymmetric Yang-Mills-Chern-Simons theory  
with the Lagrangian 
 \be
 \label{LN1}
  {\cal L} \ =\ \frac 1{g^2}  \left \langle - \frac 12 F_{\mu\nu}^2 +
  i\bar \lambda /\!\!\!\!D \lambda \right \rangle +
  \kappa   \left\langle \epsilon^{\mu\nu\rho}
  \left( A_\mu \partial_\nu A_\rho - \frac {2i}3 A_\mu A_\nu A_\rho \right ) - 
\bar \lambda \lambda \right \rangle \, .
   \ee
The conventions are:  $\epsilon^{012} = 1, \ D_\mu {\cal O}  = \partial_\mu {\cal O}  - 
i[A_\mu, {\cal O}] $ (such that $A_\mu$ is Hermitian),  
   $\lambda_\alpha$ is a 2-component Majorana $3d$ spinor belonging to the adjoint 
representation of the gauge group, and $\langle \ldots \rangle$ stands for the color trace.
   We choose
   \be
   \label{gamdef}
   \gamma^0 \ =\ \sigma^2,\ \ \ \gamma^1 = i\sigma^1,\ \ \ \gamma^2 = i\sigma^3 \ .
    \ee
 This is  a $3d$ theory and the gauge coupling constant $g^2$ carries the dimension of mass. The physical 
boson and fermion degrees of freedom in this theory are massive,
  \be
  \label{mass}
  m = \kappa g^2\ .
   \ee
 In three dimensions, a nonzero mass brings about parity breaking.  
The requirement for $e^{iS}$ to be invariant under certain large 
(noncontractible)
gauge transformations (see e.g. Ref.\cite{Dunne} for a nice review)
 leads to the quantization condition
  \be
    \label{quantkap}
    \kappa =  \frac k {4\pi}  \ .
    \ee
with integer or sometimes (see below) half-integer {\it level} $k$. 

The index \p{Witind} was evaluated in \cite{Wit99} with the result
 \be
\label{IkN}
I(k,N) \ =\  
[{\rm sgn}(k)]^{N-1} \left( \begin{array}{c} |k|+N/2 -1 \\ N-1 \end{array} \right)\ .
 \ee
for $SU(N)$ gauge group. This is valid for $|k| \geq N/2$. For $|k| < N/2$, the index vanishes
and supersymmetry is broken. In the simplest $SU(2)$ case, the index is just
 \be
\label{Ik2}
I(k,2) \ =\ k \ .
 \ee
For $SU(3)$, it is 
\be
\label{Ik3}
I(k,3) \ =\ \frac {k^2 - 1/4}2 \, .
 \ee
We can  now notice that, for the index to be integer, the level $k$ should be a half-integer 
rather than an integer
for $SU(3)$ and for all unitary groups with odd $N$. The explanation  
is that in these cases, 
the large gauge transformation mentioned above not only shifts the classical action, but also 
contributes the extra 
factor $(-1)$ due to the modification of the fermion determinant \cite{Niemi,Redlich}.

The result (\ref{IkN}) was derived in \cite{Wit99} by the following reasoning. Consider the theory in a {\it large}
spatial volume, $g^2L \gg 1$. Consider then the functional integral for the index (\ref{Witind}) 
and mentally perform a Gaussian integration over fermionic variables. 
This gives an effective bosonic action
that involves the CS term, the Yang-Mills term and other higher-derivative gauge-invariant terms. 
After that, the coefficient of the CS term is renormalized 
\footnote{This is for $k >0$. In what follows, $k$ will be assumed to be positive by default,
 although the results 
for negative $k$ are also mentioned.},
  \be
\label{kren}
 k \ \to \ k- \frac N2 \, .
 \ee

  At large $\beta$, the sum \p{Witind}  is saturated by the vacuum states of the 
theory and hence depends on 
the low-energy dynamics of the corresponding effective Hamiltonian. 
The vacuum states are determined by the term
with the lowest number of derivatives, i.e. the Chern-Simons term; the effects due to the YM term and
still higher derivative terms are suppressed at small energies and a large spatial volume. 
Basically, the spectrum of vacuum states
coincides with the full spectrum in the topological pure CS theory. The latter was determined some time 
ago 
 \begin{itemize}
\item by establishing a relationship between the pure $3d$ CS theories and $2d$ WZNW theories 
\cite{WitCMP} 
 \item  by canonical quantization of the CS theory and direct determination of the wave functions annihilated
by the Gauss law constraints \cite{Eli,Laba}.
 \end{itemize}

 The index (\ref{IkN}) is then determined as the number of states in pure CS theory 
with the shift (\ref{kren}).  For example, in the $SU(2)$ case, the number of CS states is $k+1$, 
which gives (\ref{Ik2}) after the shift.  

In Sections 2 and 3, we will rederive the result \p{IkN} using another method. We choose the spatial box 
to be {\it small} rather than large, $g^2L \ll 1$, and study the dynamics of the 
corresponding BO Hamiltonian. 
This method
was developped in \cite{Wit82} and applied there to $4d$ SYM theories. We now explain how it works.
 
Take the simplest $SU(2)$ theory.
With {\it periodic} boundary conditions for all fields 
\footnote{We stick to this choice here, although 
in a theory 
involving only adjoint fields, one could also impose  so called 
{\it twisted} boundary conditions. 
In $4d$ theories, this gives the same value for the index \cite{Wit82},  
but in $3d$ theories, the result turns out to be 
different \cite{Henningson}.}, 
the slow variables in the effective BO Hamiltonian are just the zero Fourier modes
of the spatial components of the Abelian vector potential and its superpartners,
 \be
\label{Cj}
C_j \ =\ A_j^{({\bf 0}) 3}, \ \ \ \ \ \ \ \ \ \  \lambda_\alpha \ =\ 
\lambda_\alpha^{({\bf 0}) 3} \, . 
 \ee
(In the $4d$ case, the spatial index $j$ takes three values, $j=1,2,3$;
 $\lambda_\alpha$ is the Weyl 2-component 
spinor describing
the gluino field.). The motion in the field space $\{C_j\}$ is actually finite because the shift 
  \be
\label{shift}
C_j \to C_j + 4\pi n_j/L
 \ee
with an integer $n_j$
 amounts to a {\it contractible} (this is a non-Abelian specifics) gauge transformation, under 
 which the wave functions are invariant. To the leading BO order, 
the effective Hamiltonian is nothing but the Laplacian
  \be
\label{Heff4d}
H^{\rm eff} \ =\ \frac {g^2}{2L^2}  P_j^2 \, ,
 \ee
 where $P_j$ is the momentum conjugate to $C_j$. The vacuum wave function is thus just a constant which can be multiplied by
a function of holomorphic fermionic variables $\lambda_\alpha$. We seem to have obtained  four 
vacuum wave functions of fermion charges 
0,1, and 2:
 \be
\label{4Psi}
\Psi^{F=0} \ =\ 1\, , \ \ \ \ \ \Psi^{F=1}_\alpha \ =\ \lambda_\alpha \, , \ \ \ \ \ \Psi^{F=2} = 
\epsilon^{\alpha\beta} \lambda_\alpha \lambda_\beta \, .
 \ee
However, the fermion wave functions are not allowed in this case. The matter is that the wave functions in the original theory
should be invariant under gauge transformations. For the effective wave functions, this translates into invariance
under {\it Weyl reflections}. In the $SU(2)$ case, these are just a sign flip,
 \be
\label{flip}
C_j \to -C_j, \ \ \ \ \ \ \ \lambda_\alpha \to - \lambda_\alpha\, .
 \ee
The functions $\Psi^{F=1}$ in \p{4Psi} are not invariant under \p{flip} and therefore are 
 not allowed. We are left with 2 bosonic
vacuum functions giving the value $I = 2$ for the index. A somewhat more complicated analysis (which
 is especially nontrivial for
orthogonal and exceptional groups \cite{jaKac,Keur1,Keur2}) allows  evaluating the index  for other groups. 
It coincides with the adjoint Casimir 
eigenvalue $c_V$ (another name for it is the dual Coxeter number $h^\vee$). For $SU(N)$, $I = N$.

The analysis  of the $3d$ SYMCS theories along the same lines turns out to be more complicated:
 \begin{itemize}
\item the tree level effective Hamiltonian is not just a free Laplacian, but involves an extra homogeneous magnetic field;
\item the effective wave functions are not  invariant with respect to the shifts \p{shift}, but are multiplied by certain
phase factors \cite{DJT};
 \item it is {\it not} enough to analyze the effective Hamiltonian to the leading BO order, but one-loop corrections should also be
taken into account.
  \end{itemize}
In Section 2, we will perform an accurate BO analysis at the tree level. In Section 3, 
we discuss the loop corrections.
Section 4 is devoted to the SYMCS theories with matter. 
We discuss both ${\cal N} = 1$ theories and ${\cal N}= 2$ theories. For the latter, we reproduce the results of 
\cite{ISnew}, but derive them in a more transparent and simple way. 

\section{Pure ${\cal N} =1$ SYMCS theory: the leading BO analysis.}
\setcounter{equation}0

\subsection{$SU(2)$.}

We consider $SU(2)$ theory first. As was explained above, we impose the periodic boundary conditions on all fields.
In the $3d$ case, we are left with two bosonic slow variables $C_{j=1,2} = A_j^{({\bf 0})3}$ 
and one holomorphic fermion slow variable $\lambda = \lambda^{({\bf 0})3}_1 - i\lambda^{({\bf 0})3}_2$. 
The tree-level effective BO supercharges and  Hamiltonian describe
 the motion in a homogeneous magnetic field proportional to the Chern-Simons coupling and take the form
 \be
\label{Qeff}
Q^{\rm eff}  &=& \frac {g}{L} \lambda (P_- + {\cal A}_- ) \nn
 \bar Q^{\rm eff}  &=& \frac {g}{L} \bar\lambda (P_+ + {\cal A}_+ )\, ,
 \ee
\be
\label{Heff}
H^{\rm eff} \ =\ \frac {g^2}{2L^2} \left[ \left( P_j + {\cal A}_j \right)^2 + {\cal B}
(\lambda \bar\lambda - \bar \lambda \lambda) \right] \, ,
 \ee
where 
 \be
\label{calAj}
 {\cal A}_j = \ - \frac {\kappa L^2}{2} \epsilon_{jk} C_k \, ,
 \ee
$ P_\pm = P_1 \pm i P_2$, ${\cal A}_\pm = {\cal A}_1 \pm i {\cal A}_2$, 
and ${\cal B} = \frac {\partial {\cal A}_2} {\partial C_1}   - \frac {\partial {\cal A}_1} {\partial C_2}$. 

The effective vector potential \p{calAj}
depends on the field variables $\{C_1, C_2\}$ and has nothing to do, of course, with $A^a_j(\vec{x})$. It  is defined up to a gauge transformation 
\be
\label{gauge}
{\cal A}_j \ \to \ {\cal A}_j + \partial_j f(\vec{C})  \, .
 \ee
Indeed, the particular form \p{calAj} follows from the CS terms $\sim \epsilon_{jk} A_j \dot{A}_k$ in the Lagrangian \p{LN1}, but one can always add a total time derivative to the Lagrangian, which adds a gradient to the canonical momentum $P_j$ and to the effective vector potential.

Similarly to what we had in the $4d$ case, the motion in the space $\{C_1, C_2\}$ is finite. 
However, as was  already mentioned, 
the wave functions are not
 invariant under the shifts along the cycles of the dual torus, but acquire extra phase factors,
  \be
\label{bc}
\Psi(X+1,Y) &=& e^{-2\pi i kY} \Psi(X,Y) \, , \nonumber \\
\Psi(X,Y+1) &=& e^{2\pi i kX} \Psi(X,Y) \, ,
  \ee
 where $X = C_1 L/(4\pi)$ and   $Y = C_2 L/ (4\pi) $. 
 
We explain where these factors come from. As was mentioned, the shifts $ X \to X+1$ and 
$Y \to Y+1$ represent  contractible  gauge transformations.
 In the $4d$ theories, wave functions are invariant under such transformations. 
But the YMCS  theory is special in this respect.
Indeed, the Gauss law constraint in the YMCS theory (and in SYMCS theories) is not just $D_j \Pi_j^a$, but has the form
 $$ G^a = \frac {\delta {\cal L}}{\delta A_0^a} = D_j \Pi_j^a + \frac \kappa 2 \epsilon_{jk}
 \partial_j A^a_k \ , $$
 where $\Pi_j^a = F_{0j}^a/g^2 + (\kappa/2) \epsilon_{jk}
  A^a_k $ are the canonical momenta. The second term gives rise to the phase factor associated 
with an infinitesimal gauge transformation $\delta A^a_j (\vec{x})  = D_j \alpha^a (\vec{x})$ 
(the  spatial coordinates $\vec{x}$ are not to be confused with the rescaled vector potentials $X,Y$), 
 \be
\label{fazashift}   
 \Psi[A^a_j + D_j \alpha^a] \ =  \ 
\exp\left\{-\frac {i\kappa}2 \int d\vec{x} \, \epsilon_{kl} \, \partial_k 
\alpha^a A^a_l \right \} \Psi[A^a_j] \ . 
  \ee
This property holds also for the finite contractible gauge transformations 
$\alpha^a =  (4\pi x/L) \delta^{a3}$  or $\alpha^a =  (4\pi y/L) \delta^{a3}$ 
implementing the shifts $C_{1,2} \to C_{1,2} + 4\pi/L$. 
The phase factors ${\cal E}_{1,2}(X,Y)$ thus obtained coincide with those  in  Eq. \p{bc}; they 
 are nothing but the holonomies ${\cal E}_1 = \exp \left\{i \int_{\gamma_1} {\cal A}_1 dC_1 \right\}$ and
 ${\cal E}_2 = \exp \left\{i\int_{\gamma_2} {\cal A}_2 dC_2\right\}$, 
with $\gamma_{1,2}$ being two cycles of the torus attached to the point
$(X,Y)$.  The factors ${\cal E}_{1,2}$ satisfy the property 
 \be
\label{EEEE}
{\cal E}_{1}(X,Y) {\cal E}_{2}(X+1,Y) {\cal E}_{1}^{-1}(X,Y+1) {\cal E}_{2}^{-1}(X,Y) \ = \ e^{4\pi ik} = 1 \, .
 \ee
 The phase $4\pi k$ that one acquires going around the sequence of two direct and two inverse
cycles is nothing that $2\pi \Phi$, with  $\Phi$  being the magnetic flux. For the wave functions to be uniquely defined, the latter
must be quantized. 

We note that if  another gauge for ${\cal A}_j$ were chosen, 
the holonomies ${\cal E}_{1,2}$ would be different, but the property
\p{EEEE} would of course be preserved. 

The eigenfunctions of the Hamiltonian \p{Heff} satisfying the boundary conditions \p{bc} are given by elliptic
 functions --- a variety of theta functions. There are $2k$ ground state wave functions. For $k>0$, their explicit form is 
  \be
\label{Psimtree}
  \Psi^{\rm eff}_{\rm tree}(X,Y) \ \propto  e^{-\pi k \bar z z} e^{\pi k \bar z^2}  Q^{2k}_m(\bar z),
 \ee
where $z = X + iY$, $m = 0,\ldots, 2k-1$, and the functions $Q^q_m$ are defined in the Appendix. For negative $k$, 
the functions have the same form, but with  $z$ and $\bar z$  interchanged and with 
the extra fermionic factor $\lambda$.

The index $I=2k$ of the effective Hamiltonian \p{Heff} coincides with the flux of the effective magnetic field on the dual
torus divided by $2\pi$ \cite{Novikov1,Novikov2}. 

 We next note now that not all $2|k|$ states are admissible. 
We have to impose the additional Weyl invariance condition (following
from the gauge invariance of the original theory). For $SU(2)$, this amounts to 
$\Psi^{\rm eff}(-C_j) = 
 \Psi^{\rm eff}(C_j )$ \footnote{In constrast to what should be done in 4 dimensions,  
we did not include here the Weyl reflection
of the fermion factor $\lambda$ entering the effective wave function for negative $k$. The reason is that for negative $k$,  the conveniently 
defined {\it fast} wave function (by which the effective wave function depending only on $C_j$ and $\lambda$ 
should be multiplied) 
involves  the Weyl-odd factor 
$C_1+iC_2$. This oddness compensates the oddness of the factor $\lambda$ in the effective
wave function \cite{ja1}.}  , which singles out $|k| + 1$ vacuum states, bosonic for $k > 0$ and fermionic for $k < 0$.

When $k=0$, the effective Hamiltonian \p{Heff} describes free motion on the dual torus. There are two zero energy 
ground states, 
$\Psi^{\rm eff} = {\rm const}$ and   
$\Psi^{\rm eff} = {\rm const}\cdot \lambda$ 
(we need not  bother about the Weyl oddness of the factor $\lambda$ 
by the same reason as above). The index is zero. We thus derive
 \be
\label{Itree}
I^{\rm tree}_{SU(2)} \ =\ (|k| + 1){\rm sgn} (k) \, .
 \ee  

\subsection{ Higher-rank unitary groups.}

\begin{figure} [t]
\begin{center}
\includegraphics[width=3in]{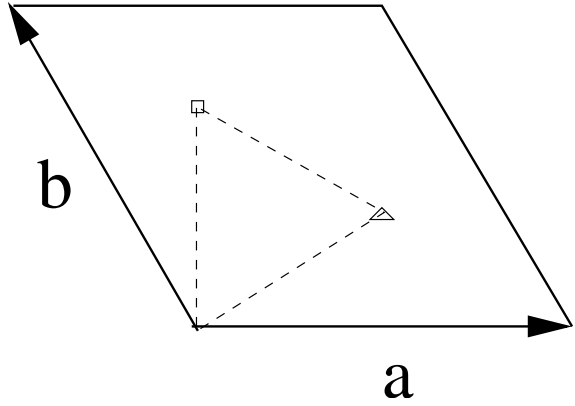}
\end{center}
\caption{Maximal torus and Weyl alcove for $SU(3)$. $\vecb{a}$ and $\vecb{b}$ are the simple coroots. 
The points $\Box$ and $\triangle$ are fundamental coweights.}
\label{romb}
\end{figure}

The effective Hamiltonian for the group $SU(N)$ involves $2r = 2(N-1)$ slow bosonic and $r = N-1$ slow fermionic 
variables $\{C^a_j, \lambda^a\}$ belonging to the Cartan subalgebra of $su(N)$ ($r$ is the rank of the group). It has the form
 \be
    \label{HeffN}
    H \ =\ \frac {g^2}{2L^2} \left[  {(P_{j}^a + {\cal A}_{j}^a)^2} +  {\cal B}^{ab}
 (\lambda^a \bar\lambda^b - \bar\lambda^b \lambda^a) \right] \ ,
    \ee
where 
\be
     \label{postojan}
     {\cal A}_{j}^a &=&   -\frac {\kappa L^2}2 \epsilon_{jk} C_{k}^a \ , \nonumber \\
     {\cal B}^{ab} &=& \kappa L^2 \delta^{ab} \ ,
    \ee
$a = 1,\ldots, r$. 
 By the same token as in the $SU(2)$ case, the motion is finite and extends over the manifold $T \times T$, with $T$
being the maximal torus of the group.
 For $SU(3)$, the latter 
is depicted  in Fig.~\ref{romb}. Each point  in Fig.~\ref{romb} is a {\it coweight} $\{w^3, w^8\}$  
such that the group element mapped on the maximal torus is  $g^{\rm torus} = \exp\{4\pi i(w^3 t^3 + w^8 t^8)\}$. The meaning of 
 the dashed lines and of special points  marked by the box and triangle is to be  explained shortly.

    The index of the effective Hamiltonian can be evaluated semiclassically by reducing the functional integral for \p{Witind} to an ordinary one  \cite{Cecotti}. The latter represents  a {\it generalized} magnetic flux
(this is nothing but that the r-{\rm th} Chern class of the $U(1)$ bundle over $T \times T$ with 
the connection ${\cal A}_{j}^a$),
   \be
    \label{indCecotti3}
    I =  \frac 1{(2\pi)^r} \int_{T \times T} \prod_{ja} dC_{j}^a \ {\rm det} \| {\cal B}^{ab} \| \ .
     \ee
    For $SU(N)$,
    \be
    \label{ISUN}
    I^{SU(N)} \ =\ N k^{N-1}\ .
     \ee

\begin{figure}[t]
\begin{center}
\includegraphics[width=1.8in]{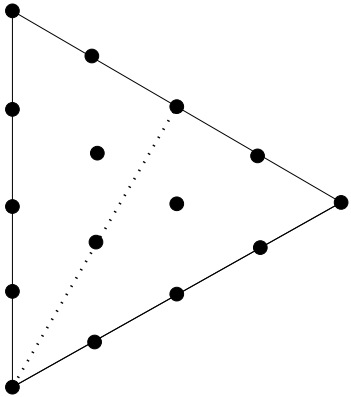}
\end{center}
\caption{$SU(3)$: 15 vacuum states for $k=4$. The dotted line marks the boundary of the Weyl alcove for $G_2$.}
\label{4k15}
\end{figure}

We find the explicit expressions for the $3k^2$ ground state wave functions in the case of $SU(3)$.
 They are given by generalized theta functions
 defined on the coroot lattice of $SU(3)$. 
They satisfy the boundary conditions
     \be
     \label{bc3}
     \Psi(\vecb{X}+\vecb{a}, \vecb{Y}) &=& e^{-2\pi i k\vecb{a} \vecb{Y}} \Psi(\vecb{X},\vecb{Y}) \ , \nonumber \\
       \Psi(\vecb{X}+\vecb{b}, \vecb{Y}) &=& e^{-2\pi ik \vecb{b} \vecb{Y}} \Psi(\vecb{X},\vecb{Y}) \ ,
       \nonumber \\
     \Psi(\vecb{X}, \vecb{Y}+\vecb{a}) &=& e^{2\pi i k \vecb{a} \vecb{X}} \Psi(\vecb{X},\vecb{Y}) \ ,
        \nonumber \\
      \Psi(\vecb{X}, \vecb{Y}+\vecb{b}) &=& e^{2\pi i k \vecb{b} \vecb{X}} \Psi(\vecb{X},\vecb{Y}) \ ,
      \ee
  where    $\vecb{X} = 4\pi \vecb{C}_1/L$,   $\vecb{Y} = 4\pi \vecb{C}_2/L$ and 
 $\vecb{a} = (1,0), \ \vecb{b} = (-1/2, \sqrt{3}/2)$ are the simple coroots.
 When  $k = 1$, there are 3 such states:
  \be
  \label{psi3k1}
  \Psi_0 &=& \sum_{\vecb{n}} \exp \left\{ -2\pi (\vecb{n} + \vecb{Y})^2 -
 2\pi i \vecb{X} \vecb{Y} - 4\pi i \vecb{X} \vecb{n} \right \} \ , \nonumber   \\
  \Psi_\triangle &=& \sum_{\vecb{n}} \exp \left\{ -2\pi (\vecb{n} + \vecb{Y} +
  {\triangle\!\!\!\!\!\triangle})^2 - 2\pi i \vecb{X}
  \vecb{Y} - 4\pi i \vecb{X} (\vecb{n} + {\triangle\!\!\!\!\!\triangle}) \right \} \ , \nonumber    \\                    
      \Psi_\Box &=& \sum_{\vecb{n}} \exp \left\{ -2\pi (\vecb{n} + \vecb{Y} +
    {\Box\!\!\!\!\!\Box})^2 - 2\pi i \vecb{X}
        \vecb{Y} - 4\pi i \vecb{X} (\vecb{n} + {\Box\!\!\!\!\!\Box} )\right \}\ ,           
   \ee
 where the sums range over the coroot lattice,  $\vecb{n} = m_a \vecb{a} + m_b \vecb{b}$
 with integer $m_{a,b}$.
 Here, ${\triangle\!\!\!\!\!\triangle}, \ {\Box\!\!\!\!\!\Box}$ are certain special points on the maximal torus
  (  fundamental coweights), such that
  $${\triangle\!\!\!\!\!\triangle} \vecb{a} =
  {\Box\!\!\!\!\!\Box} \vecb{b} = 1/2, \ \ \ \   {\Box\!\!\!\!\!\Box} \vecb{a} =
                                      {\triangle\!\!\!\!\!\triangle} \vecb{b} = 0 \ .$$
The group elements that correspond to the points $0, \triangle$, and $\Box$ belong to the center of the group, 
 \be
\label{center}
U_0 &=& {\rm diag} (1,1,1) \ , \nonumber \\
 U_\Box &=& {\rm diag}  (e^{2i\pi/3}, e^{2i\pi/3}, e^{2i\pi/3})\ , \nonumber \\
 U_\triangle &=& {\rm diag}(e^{4i\pi/3}, e^{4i\pi/3}, e^{4i\pi/3}) \ .
 \ee
They are obviously invariant with respect to Weyl symmetry, which permutes the eigenvalues.
\footnote{\label{granicy} For a generic coweight, the Weyl group elements permuting the eigenvalues $1 \leftrightarrow 2$, $1 \leftrightarrow3$ and
$2 \leftrightarrow 3$ act as the reflections with respect to the dashed lines bounding the Weyl 
alcove (the quotient $T/W$) in Fig. \ref{romb}.} 
Thus, all three states (\ref{psi3k1}) at the level $k=1$ are Weyl invariant. But for  
 $k > 1$, the number of invariant states is less than $3k^2$.  
For an arbitrary $k$, the wave functions of all $3k^2$ eigenstates
can be written in the same way as in (\ref{psi3k1}),
   \be
   \label{psi3anyk}
\Psi_n \ =\ \sum_{\vecb{n}} \exp \left\{ -2\pi (\vecb{n} + \vecb{Y} +
    \vecb{w}_n)^2 - 2\pi i \vecb{X}
        \vecb{Y} - 4\pi i \vecb{X} (\vecb{n} + \vecb{w}_n )\right \}\ , 
   \ee
where $\vecb{w}_n$ are coweights whose projections on the simple coroots $\vecb{a}, \ \vecb{b}$ represent
 integer multiples of $1/(2k)$. Only the functions (\ref{psi3anyk}) with $\vecb{w}_n$ lying in the vertices of the Weyl alcove
are Weyl invariant. For all  other $\vecb{w}_n$, one should construct Weyl invariant combinations
   \be
\label{sumWeyl}
\Psi \ =\ \sum_{\hat{x} \in W} \hat{x} \Psi_{\vecb{w}_n} \ .
 \ee 
 As a result, the number of Weyl invariant states is equal to the number 
of the coweights $\vecb{w}_n$ lying within the Weyl alcove (including the boundaries).
   For example, in the case $k=4$, there are
     15 such coweights shown in Fig.~\ref{4k15} and, accordingly, 
15 vacuum states.

       For a generic $k$, the number of the states is
 \be
I^{\rm tree}_{SU(3)}(k>0) \ =\  \sum_{m=1}^{k+1} m     \ =\ \frac {(k+1)(k+2)}2  \ .
  \ee

  The analysis for  $SU(4)$ is similar. The Weyl alcove is the tetrahedron  with the vertices
corresponding to cenral elements of $SU(4)$. A pure geometric computation gives 
                  \be
               I^{\rm tree}_{SU(4)}(k>0) \ =\  \sum_{m=1}^{k+1} \sum_{p=1}^m p   \ =\ \frac {(k+1)(k+2)(k+3)}6  \ .
                   \ee
   The generalization for  an arbitrary $N$ is obvious. It gives the result 
\be
\label{nashindextree}
I^{\rm tree}(k,N) \ =\ \left( \begin{array}{c} k+N -1 \\ N-1 \end{array} \right) \ 
 \ee

\begin{figure}[t]
\begin{center}
\includegraphics[width=4in]{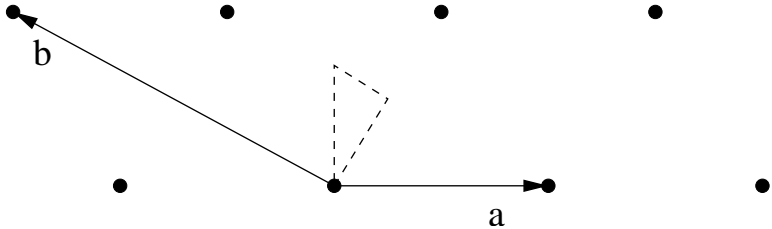}
\end{center}
\caption{Coroot lattice and Weyl alcove for $G_2$.}
\label{G2stuff}
\end{figure}

We also performed a similar analysis for the symplectic groups and for $G_2$. Let us dwell on $G_2$.
The simple coroots for $G_2$ are $\vecb{a} = (1,0)$ and $\vecb{b} = (-3/2, \sqrt{3}/2)$. The lattice of coroots
and the maximal torus look exactly in the same way as for $SU(3)$ (Fig.~\ref{G2stuff}). Hence, before Weyl-invariance 
requirement is imposed, the index
is equal to $3k^2$, as for $SU(3)$. The difference
is that the Weyl group involves now 12 rather than 6 elements, and the Weyl alcove is half size of that
 for
$SU(3)$.   As a result, for $k=4$, we have only 9 (rather than 15) Weyl-invariant states (see Fig.\ref{4k15}).
The general formula is
 \be
 \label{indG2}
 I^{\rm tree}_{G_2}( k) \ =\ \left\{ 
                 \begin{array}{c}  \frac {(|k|+2)^2}4\  \ \ \ \ \ \ \ \ {\rm for\ even} \  k \\
                                   \frac {(|k|+1)(|k|+3)}4 \ \ \ \ \ \ \ \ {\rm for\ odd} \  k 
 \end{array}
               \right\} \ .
\ee

\section{Loop corrections.}
\setcounter{equation}0

 We will mostly discuss in this section  the $SU(2)$ theory. 
For a generalization of all arguments to higher-rank groups, we refer
the reader to \cite{ja4}.
  
\subsection{Infinite volume.}
 
\begin{figure}[t]
\begin{center}
\includegraphics[width=1.3cm]{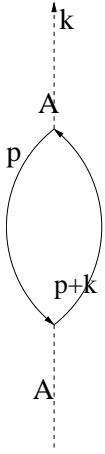}
\end{center}
\caption{Renormalization of the structure $\propto \epsilon_{\mu\nu\rho} 
 {\rm Tr} \{A_\mu \partial_\nu A_\rho\} $ by a
 fermion loop.}
\label{dvuugol}
\end{figure}

It has been known since \cite{Rao} that the CS coupling $\kappa$ in the pure YMCS theory is renormalized at the 1-loop level. 
For ${\cal N} = 1,2,3$ \ 
  SYMCS theories, the corresponding calculations
  have been performed in \cite{Kao}. 
The effect can be best understood by considering the fermion loop contribution 
to the renormalization of the structure $\propto A\partial A$ in the Chern-Simons term 
( Fig.\ref{dvuugol}). Recalling that $\kappa$ and $k$ are assumed to be positive by default, we obtain

 \be 
\label{delkap}
\Delta \kappa   \ =\ 
-mc_V \int  \frac {d^3p_E}{(2\pi)^3} \frac {1 }{(p_E^2+m^2)^2} \ =\ -\frac {c_V}{8\pi} \, .
 \ee
There is also a contribution coming from the gluon  loop. \footnote{On the other hand, spinless 
scalars (present in SYMCS theories 
with extended supersymmetries) do not contribute to the renormalization of $\kappa$.}
It is convenient \cite{ja2} to choose 
the Hamilton gauge $A_0 = 0$, in which case the gluon propagator
 \be
\label{Green3d}
D_{jk}^{ab}(\omega, \vec{p}) \  = \frac {ig^2 \delta^{ab} } {\omega^2 - \vec{p}^2 - m^2} \left[ \delta_{jk} - 
\frac {p_j p_k}{\omega^2} - \frac {im}\omega \epsilon_{jk} \right]
 \ee
involves only transverse degrees of freedom and there are no ghosts. 
 
 An accurate calculation gives
 \be
\label{renorm}
k_{\rm ren} \ = \ k + c_V - \frac {c_V}2 \, ,
 \ee
where the first term comes from the gluon loop and the second term from the fermion loop.

A legitimate question is whether the second and higher loops  also 
bring about a 
renormalization of the level $k$.
The answer is negative. The proof is simple.
      We consider the case  $k \gg c_V$. 
This is the perturbative regime where the loop corrections
 are ordered such that $\Delta k^{\rm (1\ loop)} \sim O(1),\ \  \Delta k^{\rm (2\ loops)} \sim O(1/k)$, etc. 
But  corrections to $k$ 
of the order $\sim 1/k$  are not allowed. To ensure
      gauge invariance, $k_{\rm ren}$ must be an integer. 
Hence, all higher loop
      contributions in $k_{\rm ren}$ must vanish, and they do.  
 
Note finally that the renormalization \p{renorm} refers to 
supersymmetric Yang-Mills-Chern-Simons theory --- dynamical theory with nontrivial
interactions. There is no such renormalization in the topological 
pure supersymmetric Chern-Simons theory where the fermions decouple.
The number of states in this theory is the same as in the pure CS theory.

\subsection{Finite volume.}
As was mentioned above, the coefficient $\kappa$ (with the factor $L^2$) has the meaning
  of the magnetic field on the dual torus for the effective finite-volume BO Hamiltonian.
  The renormalization of $\kappa$ translates into a renormalization of this magnetic field. At the tree level,
   the magnetic field was constant. The renormalized field is not constant, however, but depends
   on the slow variables $C_j$. To find this dependence, 
we have to evaluate the effective Lagrangian in the slow Abelian background $C_j(t)$. 
The effective vector potential is extracted from the
term $\sim {\cal A}_j(\vec{C}) \dot{C}_j$ in this Lagrangian. 
This term can be evaluated in the background field approach. Up to certain fine technical  points \cite{ja1,ja2} 
that we will not discuss here, the result can be obtained by   
taking the same Feynman graphs that determined 
renormalization of $\kappa$ in the infinite volume
and replacing 
   \be
   \label{pnC}
   p_j \to \frac{ 2\pi n_j}L - C_j \, , \ \ \ \ \ \ \ \ \ \ 
 \int \frac {d^2p}{(2\pi)^2}  \to  \frac 1{L^2} \sum_{n_j} 
\ee
in the spatial integrals there. The shift $-\vec{C}$ in the momentum is due to replacing the usual derivative
by the covariant one.

 For the effective vector-potentials induced by the fermion and the gluon loop, we derive
\footnote{The sums \p{DAferm} and \p{DAbos} diverge at large $\|\vec{n} \|$. 
Their exact meaning will be clarified shortly.}
  \be
  \label{DAferm}
{\cal A}_j^F  \ =\ 
 \frac {\epsilon_{jk}} 2 \sum_{n_j} \frac {\left(C - \frac {2\pi n} L \right)_k}
{\left(\vec{C} - \frac {2\pi \vec{n}}  L \right)^2} \left[ 1 -\frac {m} 
{\sqrt{ \left(\vec{C} - \frac {2\pi \vec{n}} L \right)^2 + m^2 }} \right] \  \stackrel{m \to 0}{\longrightarrow} \nn 
\frac {\epsilon_{jk} }2 \sum_{n_j} \frac {\left(C - \frac {2\pi n} L \right)_k}
{\left(\vec{C} - \frac {2\pi \vec{n}} L \right)^2}
  \ee
and 
  \be
\label{DAbos}
 {\cal A}_j^B \ =\ 
- \frac {\epsilon_{jk}}2 \sum_{n_j} \frac {\left(C - \frac {2\pi n} L \right)_k}
{\left(\vec{C} - \frac {2\pi \vec{n}} L \right)^2}  \left[ 2 -\frac {3m} 
{\sqrt{ \left(\vec{C} - \frac {2\pi \vec{n}} L \right)^2 + m^2 }} \right. \nn  
  \left. + 
\frac {m^3} {\left[ \left(\vec{C} - \frac {2\pi \vec{n}} L \right)^2 + m^2   \right]^{3/2}} \right] \ 
\stackrel{m \to 0}{\longrightarrow} \ 
 - \epsilon_{jk}  \sum_{n_j} \frac {\left(C - \frac {2\pi n} L \right)_k}
{\left(\vec{C} - \frac {2\pi \vec{n}} L \right)^2}
   \ee
The corresponding induced 
 magnetic fields are   
   \be
   \label{DBferm}
    \Delta {\cal B}^F(\vec{C}) &=& 
  - \frac {m }2 \sum_{n_j}
   \frac 1{\left[\left( \vec{C} - \frac {2\pi \vec{n}}L \right)^2 +
   m^2 \right]^{3/2}} \ , 
 \ee
and
\be
\label{DBbos} 
   \Delta {\cal B}^B(\vec{C}) &=&   \frac {3m }2 \sum_{n_j}
   \frac {\left( \vec{C} - \frac {2\pi \vec{n}}L  \right)^2} {\left[\left( \vec{C} - 
\frac {2\pi \vec{n}}L  \right)^2 +
   m^2 \right]^{5/2}} \, .
\ee
For most values of $C_j$, the corrections \p{DBferm}, \p{DBbos}  are of order $\sim mL^3 = \kappa g^2 L^3$,
 which is small compared to ${\cal B}^{\rm tree} \sim \kappa L^2$ if $g^2 L \ll 1$,
 which we assume. There are, however, four special points (the ``corners'' of the torus) 
\be 
\label{corners} 
C_j = 0, \ C_j = (2\pi/L, 0), \ 
C_j = (0, 2\pi/L), \ C_j = (2\pi/L, 2\pi/L) \, , 
 \ee
at the vicinity of which
 the loop-induced magnetic field is {\it much larger} than the tree-level magnetic field.  
This actually means that the 
 ``Abelian'' BO approximation, with an assumption that the energy scale associated with the slow variables 
$\{C_j, \lambda\}$ is small 
compared to the energy scale of the non-Abelian components and higher Fourier modes,
breaks down in this region. 

Disregarding this for a while, one can observe 
 that the loop corrections bring 
about effective {\it flux lines} similar to Abrikosov vortices located at the corners. 
The width of these vortices
is of the order of $m$.  Gluon corrections generate
the lines of unite flux $\Phi^B = \frac 1{2\pi} 
\int  \Delta {\cal B}^B(\vec{C})  \, d^2C \, =1$,  while the fermion loops generate the lines of  flux 
$\Phi^F = -1/2$. In total, we have a line of the flux
  $\Phi_{\rm line} = 1/2$ in each corner.

 Adding the induced fluxes to the tree-level flux, one obtains the total flux
 \be
\label{2k+2}
\Phi^{\rm tot} = 2k +4\cdot 1 -4\cdot \frac 12 = 2k+2
 \ee
suggesting the presence of 
$2k+2$ vacuum states in the effective BO Hamiltonian (before the Weyl invariance requirement is
imposed). 

Not all these states are admissible, however. The wave functions of {\it four} such states turn out to be singular
at the corners, and they should be dismissed.

Indeed, we find the effective wave functions of all 
$2k+2$ states in the Abelian valley far enough from the corners
\p{corners}.  The  effective vector potential corresponding to one of the loop-induced flux lines 
can be chosen in the form 
\be
\label{vortex}
{\cal A}_j \ =\  
-\frac {\epsilon_{jk} C_k}{2\vec{C}^2} F(m^2, \vec{C}^2)  \, .
 \ee
where the  core profile function  deduced from \p{DAferm} and \p{DAbos}, 
\be
\label{profile}
F(m^2, \vec{C}^2) \ =\ 1 - \frac {2m}{\sqrt{ \vec{C}^2 + m^2 }} + \frac {m^3}{\left(\vec{C}^2 + m^2\right)^{3/2}} 
  \ee
 vanishes at $C_j = 0$ and tends to 1 for large $C_j$. 
 
We consider the effective supercharge $Q^{\rm eff}$ given by \p{Qeff} at the vicinity of the origin, 
but outside the core of the vortex \p{vortex},  
\be
\label{regionC}
m \ll C_j \ll 4\pi/L \, .
 \ee
We can then set $F(m^2, \vec{C}^2) = 1$, neglect  the contribution of other
 flux lines as well as the homogeneous field contribution \p{calAj}. 
The equation $Q^{\rm eff} \chi^{\rm eff} = 0$ for 
a vacuum effective wave function acquires the form
 \be 
\label{Qeffchi}
\left( \frac \partial {\partial z} + \frac 1{4z} \right) \chi^{\rm eff} \ =\ 0 
 \ee
(we recall that $z = \frac {C_+L}{4\pi}$). Its solution is 
 \be
\label{chisol}
\chi^{\rm eff} (z,\bar z) \sim \frac {F(\bar z)} {z^{1/4}} \, .
 \ee

The effective 
wave function on the entire torus can be restored from two conditions:
\begin{itemize}
\item it must behave as in \p{chisol} at the vicinity of each corner;
 \item it must satisfy the boundary conditions
with the twist \p{EEEE} corresponding to the total flux \p{2k+2}. 
 \end{itemize}
This gives the structure
  \be
\label{koren4}
  \chi^{\rm eff}_m(z, \bar z) \ \propto \frac {Q^{2k+2}_m(\bar z)}{[\Pi(z) \Pi(\bar z) ]^{1/4}}  \, ,
 \ee
where 
\be
\label{Pi}
\Pi(\bar z) = Q^4_3(\bar z) - Q_1^4(\bar z) \, .
 \ee
 is a  $\theta$ function of level 
4 having zeros  at the corners \p{corners}.
\footnote{The function \p{Pi} is known from the studies of canonical 
quantization of pure CS theories \cite{Pi,Eli,Laba}. It also enters  relation \p{642}. }  

We can return now to the sums in \p{DAferm}, \p{DAbos}. The divergences 
can be regularized by subtracting from ${\cal A}_j$ a certain infinite
 pure gauge part $\sim \partial_j f(\vec{C})$ [as a side remark, this 
regularization breaks the apparent periodicity of \p{DAferm}, \p{DAbos}].
 After that, the massless limits of ${\cal A}_+^{F,B}$ and ${\cal A}_-^{F,B}$ 
represent meromorphic toric functions $P(\bar z), P(z)$ having simple poles at
 the corners \p{corners}. They are obviously expressed via $\Pi^{-1}(\bar z)$ and $\Pi^{-1} (z)$.
 
 The  full wave function is the product of the effective wave function \p{koren4} and the ground state wave function
of the fast Hamiltonian. Near the corner $\vecb{C} = 0$ in the region \p{regionC}, 
the latter behaves as $ \Psi^{\rm fast} \ \sim 1/\sqrt{|z |}$   [see Eq.(3.16) in Ref. \cite{ja2}], which is extended to the behavior
 \be    
 \label{psifastcorner}
 \Psi^{\rm fast} \ \sim \frac {1}{\sqrt{|\Pi(z )|}} \, . 
 \ee
in the whole Abelian valley.
Therefore, generically, the full wave function thus obtained is singular at the corners, 
 \be
\label{Abvalleysing}
\Psi^{\rm fast} \chi^{\rm eff}_m(z, \bar z) \ \sim \ \frac 1 {|\Pi(z)|} \, .
 \ee 

The singularity in ${\cal A}_j$  smears out when 
taking into account the finite core size suggesting that the singularity 
in the effective wave function  smears out too. However, we do not actually have a right
to go inside the core in the  Abelian BO framework: this approximation breaks down there,
as we mentioned. 

 An accurate  corner analysis (which is again 
a Born-Oppenheimer analysis, where we have to treat as slow {\it all} zero Fourier modes 
\footnote{To be precise, zero Fourier modes are relevant at the corner $C_j =0$. 
At the other corners in \p{corners}, the slow modes are characterized by 
$n_j = (1,0), \ n_j = (0,1)$, and  $n_j = (1,1)$.}  of the fields, both the Abelian and non-Abelian) that involves
the matching of the corner wave function to the wave function in the Abelian valley far from the corners  
was performed in \cite{ja2}. The result is rather natural. It turns out that the singularity 
{\it is not} smeared out when going into the vortex  core. In other words, 
the states whose Abelian BO wave functions 
 exhibit  a singularity at the corners in the massless limit, 
as in \p{Abvalleysing}, stay singular there in the exact analysis with finite mass. Such states
 are not admissible and should be disregarded.
 
The admissible wave functions still have the structure \p{koren4}, 
but theta functions $Q^{2k+2}_m(\bar z)$ should have zeros at the 
corners. In other words, they can be presented as $\Pi(\bar z)$ times a theta function of level $2k-2$. This gives
 \be
\label{koren34}
  \chi^{\rm eff}_m(z, \bar z) \ \propto  {Q^{2k-2}_m} (\bar z) \,   \Pi^{3/4}(\bar z) \, \Pi^{-1/4}(z)   \, ,
 \ee
The parameter $m$ takes now $2k-2$ values, which gives $2k-2$ [rather than $2k+2$ as would follow naively from \p{koren4}] 
``pre-Weyl'' vacuum states. After imposing the Weyl-invariance 
condition, we obtain $k$ states in agreement with \p{Ik2}. 

The following important remark is of order here. We have obtained $2k-2$ pre-Weyl states by selecting
$2k-2$ nonsingular states out of $2k+2$ states in Eq. \p{koren4}. 
This equation was obtained by taking both
gluon-induced and fermion-induced flux lines into account. 
However, it is possible to {\it eliminate} the gluon flux lines altogether.

In the region outside the vortex core where the BO approximation works, one can translate the effective Lagrangian analysis leading
to \p{DBferm} and \p{DBbos} to the effective Hamiltonian analysis. The induced vector-potentials are then
obtained as Pancharatnam-Berry phases \cite{Panch,Berry},
   \be
\label{Berry}
{\cal A}_j^{\rm PB} = -i \frac {\int (\Psi^{\rm fast})^* 
\frac {\partial }{\partial C_j} \Psi^{\rm fast} \, dx^{\rm fast} }
{\int (\Psi^{\rm fast})^* \Psi^{\rm fast} \, dx^{\rm fast} }
   \ee 
The potentials
leading to \p{DBferm} and \p{DBbos} correspond to a particular choice of $\Psi^{\rm fast}$.  
 
But we can as well 
  modify the definition of the
 fast wave function by mutiplying it by any function of slow variables. In particular, we can
multiply it by a factor that is singular at the origin ( the BO approximation is not applicable
there anyway) and define
 \be
\label{dvePsifast}
\tilde \Psi^{\rm fast} \ =\ \Psi^{\rm fast} \sqrt{ \frac {\Pi(\bar z)}{\Pi(z)} } \, .
 \ee
One cannot decide between $\Psi^{\rm fast}$ and  $\tilde \Psi^{\rm fast}$ in the Abelian BO framework. 
Evaluating \p{Berry} with $\tilde \Psi^{\rm fast}$ brings about an
 extra  gradient term [cf. \p{gauge}] giving a negative unit flux line in each corner. It annihilates
the fluxes induced by gluon loops. Now, the equation  
$Q^{\rm eff} \tilde \chi^{\rm eff} = 0$ reads 
\be 
\label{Qefftildechi}
\left( \frac \partial {\partial z} - \frac 1{4z} \right) \chi^{\rm eff} \ =\ 0 
 \ee
with the solution $\chi^{\rm eff} \sim z^{1/4} F(\bar z)$. Its extension to the entire torus is
\be
\label{korentilde}
  \tilde \chi^{\rm eff}_m(z, \bar z) \ \propto  {Q^{2k-2}_m} (\bar z) \, \sqrt{|\Pi(z)|}   \, ,
 \ee
When multiplying by $\tilde \Psi^{\rm fast}$, these functions 
({\it all} of which should be taken into account now) 
give {\it exactly the same} full wave functions as before. One can thus say that gluon-induced flux lines 
(more generally, any flux line with integer flux) should be disregarded in counting vacua. 
Such flux lines (kinds of Dirac strings) are simply not observable. 
On the other hand, vortices with fractional fluxes 
affect vacuum counting. Heuristically, four half-integer flux lines 
in a sense  ``disturb''  this counting making it ``more difficult'' for the toric vacuum wave functions 
to stay uniquely defined
(a single half-integer flux line would make it just impossible) such that the number of states is decreased.

For all other groups, the gluon loops should also be disregarded (which was recently proved in \cite{ja4}) 
and the index is obtained by substituting the value of $k$ renormalized by exclusively fermion loops,
$k \to k - c_V/2$  in 
the tree-level result.
\footnote{As we have seen, this renormalization should be 
understood {\it cum grano salis} as the renormalized magnetic field is
concentrated at the corners invalidating the Abelian BO approximation.} 
We arrive at the result \p{IkN} for $SU(N)$. For $G_2$, we obtain
\be
\label{IG2}
 I^{\rm SYMCS}_{{\cal N} = 1} [G_2] \ =\ 
\left\{  \begin{array}{c} \frac {k^2}4 \ \ \ \ \ \ \ \ \ {\rm for\ even\ } k \\  
\frac {k^2-1}4 \ \ \ \ \ \ \ \ \ {\rm for\ odd\ } k \end{array} \right. \ .
 \ee 
For completeness, we also give here  the result for the index for the 
symplectic groups. For positive $k$,
\be
\label{ISp2r}
 I^{\rm SYMCS}_{{\cal N} = 1} [Sp(2r)] \ =\ 
\left( \begin{array}{c} k + \frac {r-1}2 \\ r \end{array} \right) \, .
 \ee 
 For  negative $k$, the index is restored via $I(k) = (-1)^r I(-k)$.

\section{Theories with matter.}
\setcounter{equation}0

In  theories with matter, the index is modified compared to the pure SYMCS theories due to two effects:
  \begin{itemize}
\item an extra matter-induced renormalization of $k$;
\item the appearance of extra Higgs vacua due to nontrivial Yukawa interactions.
 \end{itemize}

The first effect seems to be rather transparent: extra fermion loops bring about extra renormalization. There are, however, subtleties to be
discussed later. 
As regards the extra Higgs vacua, 
their appearance is not limited to three dimensions, they also appear (and modify the index) 
 in $4d$ supersymmetric gauge theories. We discuss this first.

\subsection{$4d$ theories.}

Historically, it was argued in Ref. \cite{Wit82} 
that adding {\it nonchiral} matter to a theory does not change
the estimate for the index. Indeed, nonchiral fermions (and their scalar superpartners) 
can be given a mass. For large masses, they seem to decouple and the index seems to be the same as
in the pure SYM theory \footnote{This does not work for chiral multiplets, which
 are always massless and always 
affect the index  \cite{jachiral1,jachiral2}.}.  

However, it was realized later that, in {\it some} cases, massive matter 
{\it can} affect the index. The latter may
change when in  addition to the mass term,  Yukawa terms 
that couple different matter multiplets are added. The simplest example 
\footnote{It was very briefly considered in \cite{ISold} and analyzed in details 
in \cite{GVY}.} is the ${\cal N}= 1$ $SU(2)$ 
theory involving a couple of fundamental matter multiplets $Q^j_f$ ($j = 1,2$ being 
the color and $f=1,2$ the subflavor index; the indices are raised and lowered
with $\epsilon^{jk} = - \epsilon_{jk}$ and  $\epsilon^{fg} = - \epsilon_{fg}$) 
and an adjoint multiplet $\Phi_j^k = \Phi^a (t^a)_j^k$.

Let the tree superpotential be
  \be
\label{massYukawa}
{\cal W}^{\rm tree} = \mu \Phi^j_k 
\Phi^k_j + \frac m2  Q^j_f Q^f_j 
+ \frac h {\sqrt{2}} Q_{jf} \Phi^j_k Q^{kf}  \, ,
  \ee
where $\mu$ and $m$ are adjoint and fundamental masses, and $h$ is the Yukawa constant.

There is also the instanton-generated superpotential \cite{ADS},
 \be
\label{ADS}
{\cal W}^{\rm inst} \ =\ \frac {\Lambda^5}{V} \, ,
 \ee
where $\Lambda$ is a constant with the dimension of mass and 
$V =  Q^j_f Q^f_j/2 $ is the gauge-invariant moduli. 
Eliminating $\Phi$, we obtain the effective superpotential 
 \be
{\cal W}^{\rm eff} \ =\ mV - \frac {h^2 V^2}{4\mu} + \frac {\Lambda^5}V \, .
 \ee
The vacua are given by the solutions to the equation 
$\partial {\cal W}^{\rm eff} / \partial V = 0$. This equation is cubic, and hence there are 
{\it three} roots and {\it three} vacua.
\footnote{These three {\rm vacua} are intimately related to three {\it singularities} in
the moduli space of the associated ${\cal N} = 2$ supersymmetric theory with a single matter
hypermultiplet studied in \cite{SW94}.}

We now note that, when $h$ is very small, {\it one} of these vacua 
is characterized by a
 very large value,
$\langle V \rangle \approx 2\mu m/h^2$ (and the instanton term in the superpotential plays no role here). 
In the limit $ h \to 0$, it runs to infinity and we are left with only 
{\it two} vacua, the same number as in the pure SYM $SU(2)$ theory. 
Another way to see it is to observe that the equation
  $\partial {\cal W}^{\rm eff} / \partial V = 0$ becomes quadratic  for $h = 0$ having only two solutions.

The same phenomenon shows up in the theory with the $G_2$ gauge group studied in \cite{G2}. 
This 
theory involves three 7-plets $S^j_f$. The index of a
 pure SYM with $G_2$ group is known to coincide with the 
adjoint Casimir eigenvalue  $c_V$ 
of $G_2$. 
It is  equal to 4. 
However, if we 
include  the Yukawa term,
 \be
{\cal W}^{\rm Yukawa} \ =\ h\, \epsilon^{fgh} f^{jkl} 
S_{fj} S_{gk} S_{hl} 
  \ee
in the superpotential ($f^{jkl}$ being the Fano antisymmetric tensor), 
two new vacua appear. They tend to infinity in the limit $h \to 0$. 

The appearance of new vacua when Yukawa  terms are added should by no means come as a surprise. 
This  basically occurs because  the Yukawa term has a higher dimension than the mass term.

\subsection{ $3d$ superspace.}

We use a variant of the ${\cal N}=1$ $3d$ superspace formalism developped in \cite{Gates}.
The superspace $(x^\mu, \theta^\alpha)$ involves a real 2-component spinor $\theta^\alpha$.  Indices are
  lowered  and raised with antisymmetric $\epsilon_{\alpha\beta}, \epsilon^{\alpha\beta}$, $\theta^2 \equiv
 \theta^\alpha\theta_\alpha$, $\int \, d^2\theta \, \theta^2 = -2$.
The $3d$ $\gamma$-matrices $(\gamma^\mu)^\alpha_{\ \beta}$  chosen as in \p{gamdef} 
     satisfy the identity
 \be
\label{gamiden}
\gamma^\mu \gamma^\nu \ =\ g^{\mu\nu} + i \epsilon^{\mu\nu\rho} \gamma_\rho,  \, .
 \ee
 Note that $(\gamma^\mu)_{\alpha\beta}$ are all imaginary and symmetric. 
 
Gauge theories are described in terms of the real spinorial superfield $\Gamma_\alpha$. For non-Abelian theories, the
$\Gamma_\alpha$ are Hermitian matrices. As in $4d$, one can choose the Wess-Zumino gauge reducing the number
of components of $\Gamma_\alpha$. In this gauge,
 \be
\label{Gamalph}
 \Gamma_\alpha = i(\gamma^\mu)_{\alpha\beta} \theta^\beta A_\mu + i\theta^2 \lambda_\alpha \, ,
 \ee
The  covariant superfield strength is then 
 \be
\label{Walph}
W_\alpha \ =\ \frac 12 {\cal D}^\beta {\cal D}_\alpha \Gamma_\beta - \frac 12 [\Gamma^\beta, 
{\cal D}_\beta \Gamma_\alpha] \nn 
=  -i\lambda_\alpha + \frac 12 \epsilon^{\mu\nu\rho} F_{\mu\nu} (\gamma_\rho)_{\alpha\beta} \theta^\beta
+ \frac {i\theta^2} 2 (\gamma^\mu)^\beta_{\ \alpha} \nabla_\mu \lambda_\beta \,  , 
 \ee
In the superfield language, the Lagrangian \p{LN1} is written as
\be
\label{LN1super}
{\cal L} \ =\ \int d^2\theta  \left \langle  \frac 1{2g^2} W_\alpha W^\alpha  + \ \frac {i\kappa}2 
 \left( W_\alpha \Gamma^\alpha + \frac 13 \{\Gamma^\alpha, \Gamma^\beta\}  
{\cal D}_\beta \Gamma_\alpha 
\right) \right \rangle \, . 
 \ee
 
We now add matter multiplets. In this talk, we will consider only  real adjoint multiplets. 
(In Ref.\cite{ja3}, we also treat
 the theories with complex fundamental
multiplets.) Let there be only one such multiplet,
 \be
\label{Sigma}
\Sigma = \sigma + i\psi_\alpha \theta^\alpha + i\theta^2 D \, .
 \ee
The gauge invariant kinetic term has the form
 \be
\label{LPhikin}
{\cal L}^{\rm kin} = - \frac 1{2g^2}  \int d^2\theta  \left \langle \nabla_\alpha \Sigma \, \nabla^\alpha \Sigma  
\right\rangle\, ,
 \ee
where $\nabla_\alpha \Sigma = {\cal D}_\alpha \Sigma - [\Gamma_\alpha, \Sigma]$. One can add also the mass term
\footnote{This is a so called {\it real} mass term which can be expressed in terms of ${\cal N} =2$ $3d$ superfields only by
 adding an explicit $\theta$-dependence in the integrand  in $S = \int \, d^4 \theta \cdots$ \cite{Nishino,Ahamass,Boer}. 
In a theory with two real multiplets \p{Sigma}
(which form a chiral ${\cal N} = 2$ multiplet), one can also write down theа complex mass term in the same way as in $4d$ theories.
 Such terms do not
renormalize $k$, however,  and we will not consider them here.},
 \be 
\label{LPhiM}
{\cal L}_M \ =\ -i\zeta  \int d^2\theta \langle \Sigma^2 \rangle  \, .
 \ee

Adding 
\p{LN1super}, \p{LPhikin}, \p{LPhiM},  expressing the Lagrangian in components, and eliminating the auxiliary field
$D$, we obtain
  \be
 \label{LN2}
   {\cal L} \ =\ \frac 1{g^2}
  \left\langle - \frac 12 F_{\mu\nu}^2 + \nabla_\mu \sigma \nabla^\mu \sigma + 
   \lambda/\!\!\!\!\nabla \lambda  +  \psi/\!\!\!\!\nabla \psi \right\rangle \nonumber \\ +
  \kappa   \left\langle \epsilon^{\mu\nu\rho}
  \left( A_\mu \partial_\nu A_\rho - \frac {2i}3 A_\mu A_\nu A_\rho \right ) + i \lambda^2  \right\rangle 
 + i\zeta 
\langle \psi^2 \rangle  - \zeta^2 g^2 \langle \sigma^2 \rangle  \, .
   \ee 
 Besides the gauge field, the Lagrangian 
 involves the adjoint fermion $\lambda$ with the mass $m_\lambda = \kappa g^2$, the adjoint 
fermion  $\psi$ with the mass $m_\psi = \zeta g^2$ and the adjoint scalar $\sigma$ with the same mass. The point $\zeta = \kappa$
is special. In this case, the Lagrangian \p{LN2} enjoys  ${\cal N} = 2$ supersymmetry.

\subsection{Index calculations}

We consider  the theory defined by \p{LN2}. First, let  $\zeta > 0$. Then the mass of the 
matter fermions is positive. To be more precise, it has the same sign as the gluino mass for $k > 0$. The matter loops  lead to an extra  renormalization of $k$.

We note that the status of this renormalization is different from the one due to the gluino loop. We have seen that for the latter, the induced magnetic
field on the dual torus is concentrated at the corners \p{corners}, which follows from the equality $m_\lambda L \ll1$. 
 On the other hand, the mass of the matter fields $m_\psi = \zeta g^2$ is an independent parameter. 
It is convenient to make it {\it large}, $m_\psi L \gg 1$. For a finite mass, the induced magnetic field has the form as in Eq.\p{DBferm}.
For small $m_\psi L$, it is concentrated at the corners. But in the opposite limit, the induced flux density becomes constant, as the
tree flux density is. 

Thus, massive enough matter brings about a true renormalization of $k$ without any 
 qualifications ({\it sine sale} if you will).

For positive $\zeta$, the renormalization is negative, $k  \to  \ k-1$. The index coincides with 
the index of the ${\cal N} = 1$ SYMCS theory with a renormalized $k$,
  \be
\label{poszeta}
 I_{\zeta > 0}  = k -1 \, .
 \ee
For $k = 1$, the index is zero and supersymmetry is spontaneously broken.

For negative $\zeta$, two things happen. 

\begin{itemize}

\item
First, the fermion matter mass has the opposite sign and so does the renormalization of $k$ due
to the matter loop. We seem to obtain $I_{\zeta < 0} =  k + 1$.

\item This is wrong, however, due to another effect. For a positive $\zeta$, the  ground state wave function 
in the matter sector is bosonic. But for a negative $\zeta$, 
it is fermionic, $\Psi \propto \prod_a \psi^a$, changing the sign of the index. 

\end{itemize}

 We obtain
 \be
\label{negzeta}
 I_{\zeta < 0} = -k - 1 \, .
 \ee
Supersymmetry is broken here for $k = -1$. 

As was mentioned, the Lagrangian \p{LN2} with $\zeta = \kappa$ enjoys 
the extended ${\cal N} = 2$ supersymmetry. That means, in particular,
that $\zeta$ changes a sign together with $\kappa$ and the result 
is given by 
 \be
\label{IndN2}
I^{\rm SYMCS}_{{\cal N} = 2} \ =\ |k| - 1 
 \ee
in agreement with \cite{Ohta,Bergman}.
 In contrast to
\p{poszeta} and \p{negzeta}, this expression is not analytic at $k=0$, the nonanalyticity being  due just to  the sign flip 
 of the matter fermion mass. 

Strictly speaking, 
formula \p{IndN2} does not work for $k=0$. In this case, 
we should also keep $\zeta = 0$, 
the matter is massless, massless scalars make the
motion infinite, and the index is ill-defined. However, bearing 
in mind that the regularized theory with $\zeta \neq 0$ gives the 
result  $I^{\rm SYMCS}_{{\cal N}= 2\ {\rm deformed}}(0) = -1$, 
irrespectively of the sign of $\zeta$, one can attribute this value for the index also to
$ I^{\rm SYMCS}_{{\cal N}= 2 }(0)$. 
  
\begin{figure}[ht!]
     \begin{center}

            \includegraphics[width=0.6\textwidth]{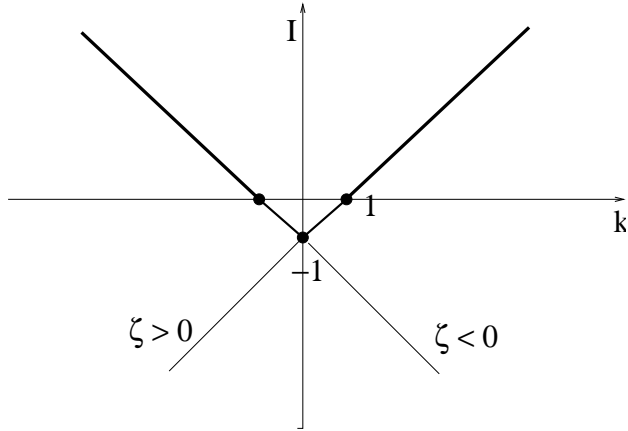}
        
    \end{center}
    \caption{ The indices in the theory 
\p{LN2} with $\zeta > 0$, $\zeta < 0$, and $\zeta = \kappa$ (bold lines). }
\label{triind}
    \end{figure}

The three index formulas \p{poszeta}, \p{negzeta}, and \p{IndN2} are represented together in Fig.~\ref{triind}.

We next consider the theory involving a gauge multiplet \p{Gamalph} 
and {\it three} ${\cal N} =1$ real adjoint matter multiplets \p{Sigma}.
If all the masses are equal, the theory has the extended ${\cal N} = 2$ supersymmetry provided an 
extra Yukawa term
is added in the Lagrangian \cite{Ivanov}. If we call one of the matter multiplets 
(the one that forms  together with \p{Gamalph} a  ${\cal N} = 2$ superfield) $\Sigma$ and combine 
two other  real multiplets into the  complex  ${\cal N} = 2$  multiplet $\Phi$, then the Yukawa term acquires the form
 \be
{\cal L}_{\rm Yukawa} \ =  -\frac {2i} {g^2} \int d^2\theta \, \langle  \Sigma  \Phi \bar \Phi \rangle \, .  
 \ee

To calculate the index, however, we will consider the {\it deformed} ${\cal N} = 1$ Lagrangian with different masses $M$ 
(the mass of the field $\Sigma$) and $m$ (the mass of the complex field $\Phi$). 
We  assume $M$ to be large, but, to make the transition to the
${\cal N} =2$ theory smooth, its sign to coincide with the sign
of  $k$.

There are four different cases:
\footnote{When comparing with \cite{ISnew}, note the mass sign convention for the matter fermions there is
 {\it opposite}  to our convention. We call the mass positive if it has the same sign as the 
masses of fermions in the gauge multiplet for positive $k$ (and hence positive $\zeta$). In other words, for 
positive $k,\xi$, the shifts of $k$
due to both gluino loop and adjoint matter fermion loop have the negative sign.}
 
\begin{enumerate}

\item $m > 0, \ \ k>0 \Rightarrow M > 0$.
\be
\label{krenSU2}
k  \to  k -  1_{\Sigma} - 2_{\Phi} = k-3 \, .
 \ee
This contributes $k -  3$ to the index. Note that, for $k=  1,2$, this contribution is negative.

\item $m > 0, \ \ k,M < 0$.
 \be
k  \to  k  + 1_{\Sigma} - 2_{\Phi} = k-1
 \ee
Multiplying it by -1 due to the fermionic nature of the wave 
function [ in this case, it involves a fermionic factor associated with the  real adjoint matter multiplet $\Sigma$;
 see the discussion before Eq.\p{negzeta}], we obtain $I = -k +1$.

\item $m < 0, \ \ k,M >0$.
 \be
 k \to  k  - 1_{\Sigma} +  2_{\Phi} = k+1 \, ,
 \ee
giving the contribution $I = k+1$.

\item $m < 0, \ \ k,M<0$.

 \be
k  \to  k  + 1_{\Sigma} + 2_{\Phi}
 = k+ 3 \, .
 \ee
The contribution to the index is $ -k - 3$.

\end{enumerate}
  
In contrast to the model with only one real adjoint multiplet, this is not the full answer yet. 
There are also additional states on the Higgs branch that contribute to the index. Indeed, the superpotential is 
 \be
\label{WSigPhi}
{\cal W} \ = \ \frac {-i}{g^2} \left( \frac M2 \Sigma^a \Sigma^a + m \bar \Phi^a \Phi^a + i \epsilon^{abc} \Sigma^a \Phi^b \bar \Phi^c 
\right)\, .
 \ee
The bosonic potential vanishes provided
 \be
\label{uslovminadj}
m\phi^b\  = \ i\epsilon^{abc} \sigma^a \phi^c \, , \nonumber \\
 M\sigma^a \ =\ i  \epsilon^{abc}  \bar \phi^b \phi^c \, .
 \ee
 These equations have  nontrivial solutions when both $M$ and $m$ are positive or when both $M$ and $m$ are negative. Let them be positive.
Then  \p{uslovminadj} has a unique solution up to a gauge rotation. 
   \be
\label{soladj}
   \sigma^a \ = \  m \left( \begin{array}{c} 0 \\ 0 \\ 1 \end{array} \right) , \ \ \ \ \ \ \phi^a = \sqrt{\frac {Mm}2} 
\left( \begin{array}{c} 1 \\ -i \\ 0 \end{array} \right)\, . 
 \ee

The corresponding contribution to the index is not just equal to 1, however, due to a new important effect
that did not take place in $4d$ theory with superpotential \p{massYukawa} 
considered above and would also be absent in a $3d$ theory with a fundamental
${\cal N} =2$ matter multiplet. 

Indeed, besides the solution \p{soladj}, 
there are also the solutions obtained
from that by  gauge transformations. The latter are not necessarily global, 
they might depend on the spatial coordinates $x,y$.
We note that, for the theory defined on a torus,  certain 
transformations  can be applied to \p{soladj} that look like gauge
transformations, but are not  contractible due to the nontrivial $\pi_1[SO(3)] = Z_2$. (Here, 
 $SO(3)$ should be understood  not as the orthogonal group itself, but rather as the  adjoint representation space; cf. the discussion
of higher isospins below.)
An example of such a quasi-gauge transformation is 
  \be
\label{largegauge}
\Omega_1: O^{ab}(x) \ =\ \left( \begin{array}{ccc} 
\cos \left( \frac {2\pi x}L \right) &  \sin \left( \frac {2\pi x}L \right) & 0 \\
-\sin \left( \frac {2\pi x}L \right) & \cos \left( \frac {2\pi x}L \right) & 0 \\
 0 & 0& 1   \end{array}   \right) \, ,
  \ee
where $L$ is the length of our box. The transformation \p{largegauge} does not affect 
$\sigma^a = \sigma \delta^{a3}$ and keeps the fields 
$\phi^a(\vec{x})$ periodic. 
\footnote{For matter in fundamental representation, the transformation 
\p{largegauge} is inadmissible:
when lifted to $SU(2)$, it would make a constant solution, the fundamental analog of \p{uslovminadj}, antiperiodic.} 
There is a similar transformation $\Omega_2$ along the second cycle of the torus.

In $4d$ theories, wave functions are invariant under contractible gauge  transformations. In $3d$ SYMCS theories, they
are invariant up to a possible phase factor, as in \p{bc}. 
But nothing dictates the behaviour of the wave functions under the transformations $\Omega_{1,2}$ which are actually {\it not}
gauge symmetries, but rather some global symmetries of the theory living on a torus. We thus obtain  four different wave 
functions, even or odd under the action of $\Omega_{1,2}$. 
\footnote{The oddness of a wave function under the transformation \p{largegauge} 
means  a nonzero {\it electric flux} in the language
of Ref. \cite{Hooft}.}  
The final result for the index of this theory is
   \be
 \label{IndN2adj2}
 I^{{\cal N} = 2}_{\rm  adjoint \ matter}  \ =\ |k| + 1 \, .
  \ee
universally for positive and negative $k$.  Extra Higgs states contribute only for positive $k$.

The result   \p{IndN2adj2}  
was  derived among others in \cite{ISnew} following a different logic. Intriligator and Seiberg 
did not deform  ${\cal N} = 2 \ \to {\cal N} = 1 $, but kept the fields in the  
real adjoint matter multiplet $\Sigma$ light. Then the light matter fields $\{ \sigma, \psi\}$ 
 enter the effective BO Hamiltonian at the same ground as the Abelian components of the gluon and gluino fields.
 As was mentioned,
the fluxes induced
by the light fields are not homogeneous being concentrated at the corners. 
This makes an accurate analysis  substantially more difficult.
The index \p{IndN2adj2} was obtained in \cite{ISnew} as a sum of 
{\it three} rather than just two contributions 
\footnote{On top of the usual vacua with $\phi = \sigma =0$ and the Higgs vacua with $\phi,\sigma \neq 0$, 
they also had  ``topological vacua'' with $\phi=0, \sigma \neq 0$. These do not appear in our approach. }
and it is still not quite clear  how this works  in the particular case $k=2$ where
$k_{\rm eff}$ as defined in  Ref.\cite{ISnew} and including only renormalizations 
due to complex matter multiplet, $k_{\rm eff} = k-2$,
vanishes. 

   Our method is simpler.

We can also add the ${\cal N}= 2$ multiplets with higher isospins. Then the counting of Higgs vacuum states becomes more complicated. For
example, for $I=3/2$, there are 10 such states. This number is obtained as a sum of the single state with the isospin projection $1/2$ and 
$3^2 = 9$ states with the isospin projection $3/2$ (in the latter case, there is a constant solution supplemented by
eight $\vec{x}$-dependent quasi-gauge copies). 
The generic result for the index in the theory involving several ${\cal N} =2$ matter multiplets with different isospins is
 \be
\label{mnogo}
I = |k| - 1 + \frac 12 \sum_f T_2(I_f) \, ,
 \ee 
where 
 \be
\label{Dynkin} 
  T_2(I) \ = \frac {2I(I+1)(2I+1)} 3 
  \ee
is the Dynkin index of the corresponding representation normalized to $T_2({\rm fund}) = 1$.

When deriving \p{mnogo}, it was assumed that the matter-induced shift of the index is the sum of the individual shifts due to individual multiplets.
This is true if the Lagrangian does not involve extra cubic ${\cal N} = 2$ invariant superpotentials which can bring about 
 extra Higgs vacuum states.

We can observe that the index does not depend on the sign of $k$, although this universal result is obtained by adding
 the contributions that look completely different for $k>0$ and $k<0$. 
For an individual multiplet contribution, the Higgs states contribute only for one sign of $k$ 
 (positive or negative depending on the sign of the mass). An interesting 
explanation of the symmetry under the  mass sign flip with a given $k$ (and hence under the sign flip of $k$ with a given $m$) 
was suggested in \cite{ISnew}. 
Basically, the authors argued that one can add to the mass the size of one 
of the dual torus  cycles times $i$ 
to obtain  a complex holomorphic parameter on which the index of an ${\cal N} = 2$ theory 
should not depend. Hence, it should not depend on the real part of this
parameter (the mass). We believe that it is still dangerous to pass the point $m=0$ where the 
index is not defined and this argument 
therefore lacks rigour. Anyway, an explicit $SU(2)$ calculation shows that the symmetry with respect to mass sign flip is  indeed
maintained.

The reasoning above can be generalized to  higher-rank unitary groups. 
Intriligator and Seiberg conjectured the following generalization of \p{mnogo},
    \be
\label{IndSUNfinal}
 I^{SU(N)} &=&  \frac 1{(N-1)!} \prod_{j = -\frac N2 +1}^{\frac N2 -1} 
\left( |k| + \frac 12  \sum_f T_2(R_f) - \frac {N} 2 -j \right) \, ,
   \ee
implying that the overall shift of $k$ is represented as the sum of individual shifts due to indivudual multiplets.
For an individual contribution to the shift, 
this formula can be derived for different signs of $k$ and $m$ when the extra Higgs states do not contribute. 
It can be extended to $k,m$ of the same sign
using the symmetry discussed above.  

We checked that this works for
 all $SU(N)$ groups with fundamental matter 
 and for  $SU(3)$  with adjoint matter.
\footnote{We emphasize that this is all 
  specific to  ${\cal N} = 2$. For ${\cal N} = 1$ theories, 
there is no such symmetry, 
 (see, e.g., Fig. 2).}
It would be interesting to construct a rigourous proof of this fact.

\section*{Appendix A. Theta functions.}
\setcounter{equation}0
\def\theequation{A.\arabic{equation}} 
\vspace{1mm}

We here recall  certain mathematical facts 
concerning the properties of analytical functions 
on a torus. They are mostly taken from the textbook \cite{Mumford}, but we are using a different notation,
which we find  clearer and more appropriate for our purposes.

Theta functions play the same role for the torus as ordinary polynomials for the Riemann sphere. 
They are analytic, but satisfy certain nontrivial  quasiperiodic boundary conditions with respect to shifts
 along the cycles of the torus. 
A generic torus is characterized by a complex modular parameter $\tau$, but we will stick to the 
simplest choice $\tau = i$ so that the torus represents a square $x,y \in [0,1]$ ( $z = x+iy$) glued around.

The simplest $\theta$-function satisfies the boundary conditions
 \be
\label{thet1bc}
 \theta(z+1) &=& \theta(z) \, ,\nonumber \\
 \theta(z+i) &=& e^{ \pi (1 - 2iz)} \theta(z) \, .
 \ee
This defines a {\it unique} (up to a constant complex factor) analytic function. Its explicit
form is
 \be
\label{thet1}
\theta(z) \ =\ \sum_{n = -\infty}^\infty \, \exp\{- \pi n^2 + 2\pi i n z \} \, .
 \ee
This function (call it theta function of level 1 and introduce an alternative
notation $\theta(z) \equiv Q^1(z)$) has only one zero in the square $x,y \in [0,1]$, 
exactly in its middle, 
$\theta(\frac {1+i}2 ) = 0$.

For any integer $q > 0$, theta functions
of level $q$  can be defined  such that
 \be
\label{thetqbc}
 Q^q(z+1) &=& Q^q(z) \, ,\nonumber \\
 Q^q(z+i) &=& e^{q \pi  (1 - 2iz)} Q^q(z) \, .
 \ee
The boundary conditions \p{thetqbc} involve the twist [the exponent in the R.H.S. of Eq. \p{EEEE}] $-2\pi q$
corresponding to the negative {\it magnetic flux}.
\footnote{This is a physical interpretation. Mathematicians would call it {\it monodromy}. } 
The functions $Q^q(\bar z)$ have  positive fluxes $2\pi q$. Multiplying 
$Q^{2k}(\bar z)$ and $Q^{2k}(z)$ by proper exponentials, 
 yields the functions \p{Psimtree} (no longer analytic) 
satisfying the boundary conditions \p{bc}.

The functions satisfying (\ref{thetqbc}) lie in a vector space of dimension $q$. The basis
in this vector space can be chosen as 
 \be
\label{Qqm}
Q^q_m(z) \ =\ \sum_{n = -\infty}^\infty \, \exp\left\{- \pi q \left(n + \frac mq \right)^2 + 2\pi i q z 
\left( n + \frac mq \right)\right\} \, , \nn
 m = 0, \ldots, q-1 \, .
 \ee
  In Mumford's notation \cite{Mumford}, $Q^q_m(z)$ can be expressed as
  \be 
\label{QviaMum}
 Q^q_m (z) \ =\ \theta_{m/q,0} (qz, iq) \ ,
 \ee
where $\theta_{a,b}(z, \tau)$ are theta functions of rational characteristics. 

$Q^q_m(z)$ can be called  ``elliptic polynomials'' of degree $q$. Indeed, each $Q^q_m(z)$ 
has $q$ simple zeros at
 \be
\label{zs}
z_s^{(m)} \ =\ \frac {2s+1}{2q} + i \left(  \frac 12   - \frac mq \right)\, , \ \ \ \ \ \ \ 
s= 0,\ldots,q-1 
 \ee
(add $i$ to bring it onto the fundamental domain $x,y \in [0,1]$ when necessary). 
A product $Q^q(z) Q^{q'}(z)$ of two such ``polynomials'' of degrees $q, q'$ gives a polynomial of degree
 $q + q'$. There are many relations between the theta functions of different level and their products, which follow. 
 We can amuse the reader with a relation
 \be
\label{642}
\frac {Q^6_5(z) - Q^6_1(z)}{(Q^4_3(z)- Q^4_1(z)) Q^2_0(z)}  \ =\ \frac 1{\eta(i)} = 
\frac {2\pi^{3/4}}{\Gamma(1/4)} \, .
 \ee
 
The ratios of different elliptic functions of the same level give double periodic meromorphic elliptic functions. For example,
the ratio of a properly chosen linear combination  $\alpha Q^2_1(z) + \beta Q^2_2(z)$ and $[\theta(z)]^2$ is the Weierstrass function.

\end{document}